\documentclass[twocolumn,prl]{revtex4}

\usepackage[dvips]{graphicx}
\usepackage{dcolumn}
\usepackage{bm}

\begin{document}

\preprint{MMM/Intermag/CR-03}

\title{
Dielectric responses of the layered cobalt oxysulfide Sr$_2$Cu$_2$CoO$_2$S$_2$
with CoO$_2$ square-planes
}

\author{S. Okada}
\email{sokat@htsc.sci.waseda.ac.jp}

\author{I. Terasaki}%
\affiliation{%
Department of Applied Physics Waseda University,
Tokyo 169-8555, Japan
}

\affiliation{%
CREST, Japan Science and Technology Agency,
Tokyo 108-0075, Japan
}%

\author{H. Ooyama}
\author{M. Matoba}
\affiliation{
Department of Applied Physics and Physico-Informatics, Keio University,
Yokohama 223-8522, Japan}%

\date{\today}

\begin{abstract}
We have studied the dielectric responses of the layered cobalt oxysulfide 
Sr$_2$Cu$_2$CoO$_2$S$_2$ with the CoO$_2$ square-planes. 
With decreasing temperature below the N\'eel temperature, 
the resistivity increases like a semiconductor,
and the thermopower decreases like a metal. 
The dielectric constant is highly dependent on temperature, and
the dielectric relaxation is systematically changed with temperature,
which is strongly correlated to the magnetic states.
These behaviors suggest that carriers distributed homogeneously 
in the paramagnetic state at high temperatures
are expelled from the antiferromagnetically ordered spin domain 
below the N\'eel temperature. 
\end{abstract}

\maketitle

\begin{figure}[t]
\includegraphics[width=.7\linewidth]{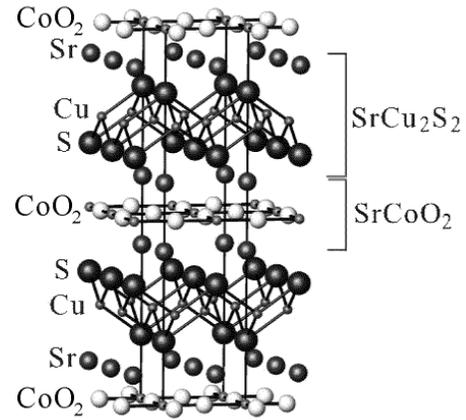}
\caption{Crystal structure of Sr$_2$Cu$_2$CoO$_2$S$_2$ 
which crystallizes in an unusual intergrowth 
structure with the stacking sequence -Sr-Cu$_2$S$_2$-Sr-CoO$_2$-Sr-
(space group $I/4mmm$)~\cite{okada02}.}
\label{struct}
\end{figure}

Many layered transition-metal oxides and sulfides 
have been the subject of the experimental and theoretical investigations, 
because they exhibit unusual physical properties originating 
from the strong electron correlation.
From the viewpoint of materials design, 
we have studied the prototypical transition-metal hybrid material
Sr$_2$Cu$_2$CoO$_2$S$_2$~\cite{chu}.
Layered cobalt oxysulfide Sr$_2$Cu$_2$CoO$_2$S$_2$ crystallizes 
in an unusual intergrowth structure 
consisting of the CoO$_2$ square lattice 
and ThCr$_2$Si$_2$-type sulfide layers, 
as shown Figure \ref{struct}. 

\begin{figure}[t]
\includegraphics[width=.7\linewidth]{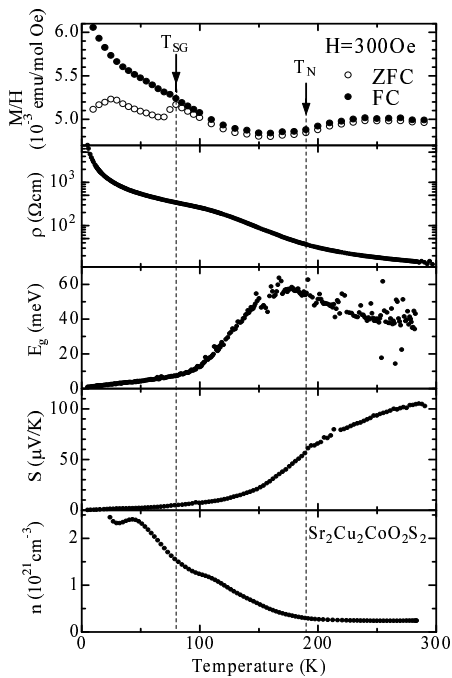}
\caption{Temperature dependences of (a) field-cooled and
zero-field-cooled magnetic susceptibilities, 
(b) electrical resistivity ($\rho$), 
(c) activation energy estimated from $\rho$, 
(d) thermopower ($S$), 
and (e) local carrier density estimated from $S$.}
\label{pp}
\end{figure}

Sr$_2$Cu$_2$CoO$_2$S$_2$ exhibits rich and complicated magnetism,
and shows successive transitions.
First, it shows an antiferromagnetic transition 
with the N\'eel temperature $T_N$=190 K 
which is detected by the neutron powder diffraction (NPD). 
$M/H$ in Figure \ref{pp} (a) 
shows a broad maximum near 250 K, being indicative of 2D antiferromagnetic nature, 
as is normally seen for antiferromagnetic K$_2$NiF$_4$-type compounds. 
Below $T_{N2}=$125 K, additional peaks appear in the NPD pattern, 
which suggests that interlayer magnetic correlation is grown. 
Below $T_{SG}=$80 K, Sr$_2$Cu$_2$CoO$_2$S$_2$ has a hysteresis behavior
between field-cooled and zero-field-cooled magnetization curves, 
as is often seen in frustrated spin systems \cite{okada01,okada02,okada03}.

In transition-metal compounds, charge and spin degrees of freedom
interact with each other to induce unconventional physics, 
such as high-temperature superconductivity in Cu oxides.
Thus, in order to explore effects of magnetic excitations on 
charge transport of Sr$_2$Cu$_2$CoO$_2$S$_2$, 
we have measured electrical resistivity ($\rho$), thermopower ($S$)
and dielectric constant ($\varepsilon$) 
of polycrystalline samples.

Polycrystalline samples of Sr$_2$Cu$_2$CoO$_2$S$_2$ were prepared 
by a solid-state reaction\cite{okada01,okada02}.

The resistivity was measured by a four-probe method, and 
the thermopower was measured using a steady-state technique
from 4.2 from 300 K in a liquid He cryostat. 
The dielectric constant ($\varepsilon _1$) and ac conductivity ($\sigma_{ac}$) 
were measured with a parallel plate capacitor arrangement 
using an ac two-contact four-probe method 
with an LCR meter (Agilent 4284A) from 10$^2$ to 10$^6$ Hz.
Owing to the two-probe configuration,
the contact resistance and capacitance might have affected the measurement.
However, the contact resistance was less than 1 $\Omega$,
which was much smaller than the sample resistance (50 $\Omega$ at 200 K).
In fact, the observed $\sigma_{ac}$ was quantitatively consistent with
the observed dc resistivity for $\omega\to 0$.
Though we did not evaluate the contact capacitance, 
we can employ an evaluated value of 500 pF for La$_2$CuO$_4$ 
from Ref. \cite{chen},
because the contact capacitance is primarily determined 
by the area of the contact.
It gives a reactance of 3$\times$10$^5~\Omega$ at 1 kHz, 
which is 10$^5$ times larger than the contact resistance.
As a result, we can safely assume that our contact is primarily 
a resistive coupling, and can neglect the contact capacitance.
We should further note that we tested different contact configurations
of the same sample, and got identical results within experimental errors.
Thus we conclude that the contact would not affect the measured $\varepsilon$
within experimental errors of 1\%.

Figure \ref{pp} (b) shows the temperature dependence of $\rho$. 
The $\rho$ increases with decreasing temperature, 
which indicates that Sr$_2$Cu$_2$CoO$_2$S$_2$ is a semiconductor.
As is clearly shown, the temperature dependence of $\rho$ is not a simple
activation type. 
In particular, $\rho$ shows a ``plateau'' near 50-120 K, 
indicating that this material is more metallic than a conventional 
semiconductor with a constant activation energy.
In order to see this more clearly, 
the activation energy ($E_g$) estimated from the relation as
\begin{eqnarray}
\rho (T)=\rho _0 \exp({E_g}/{k_BT}),
\end{eqnarray}
is plotted as a function of temperature in Fig. \ref{pp} (c).
As expected, $E_g$ depends on temperature. 
In particular, $E_g$ suddenly decreases below around $T_N$, and 
is as small as $E_g\sim k_BT$ below around $T_{SG}$.
This suggests the carriers feel essentially
no energy gap, and can move more freely at low temperature. 
In this sense, the electronic state below $T_{SG}$
is ``metallic'' rather than semiconductive. 

Figure \ref{pp} (d) shows the temperature dependence of $S$. 
$S$ decreases with decreasing temperature, 
and $S/T$ is as small as $S/T$ of conventional metals below around $T_{SG}$.
This seems to indicate that Sr$_2$Cu$_2$CoO$_2$S$_2$ is a metal
at low temperatures,
and is consistent with the anomalously small $E_g$.
In the lowest order approximation, 
$S$ is expressed as functions of the carrier concentration ($n$) as 
\begin{eqnarray}
|S| =\frac{\pi k_B^2}{2 \hbar ^2 d_c e}\frac{m}{n}T,
\end{eqnarray}
where $d_c$ (=17.7 \AA) is the inter-layer spacing, 
$e$ ($> 0$) is the unit charge, and 
$m$ is the bare electron mass \cite{S2D}.
Fig. \ref{pp} (e) shows the temperature dependence of $n$.
Quite anomalously, $n$ \textit{increases} 
with decreasing temperature below $T_N$, which could not happen in 
a homogeneous system.

Suppose that the carriers (holes) and magnetic spins are 
segregated in the sample. 
Then the electric current flows in a filamentary path 
of the hole-rich region connected in a percolation network across the sample.
In this situation,  $n$ observed from $S$ should be the carrier density 
in the filamentary path.
If the volume fraction of the filaments are 10 \%, $n$ would be 10 times
larger than $n$ distributed homogeneously.
The large $n$ and the small $E_g$ naturally suggest
that the carriers are segregated to form a percolation network below $T_N$.

\begin{figure}[t]
\includegraphics[width=.75\linewidth]{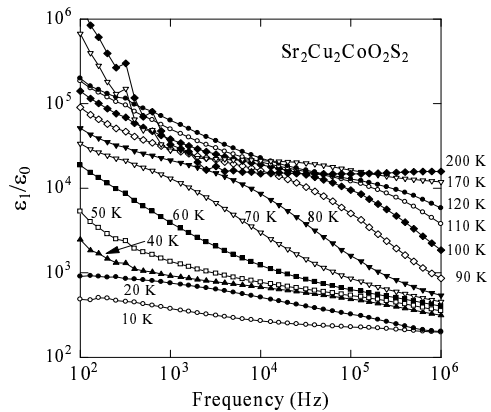}
\caption{Frequency dependence of dielectric constant. }
\label{f3}
\end{figure}

\begin{figure}[t]
\includegraphics[width=.65\linewidth]{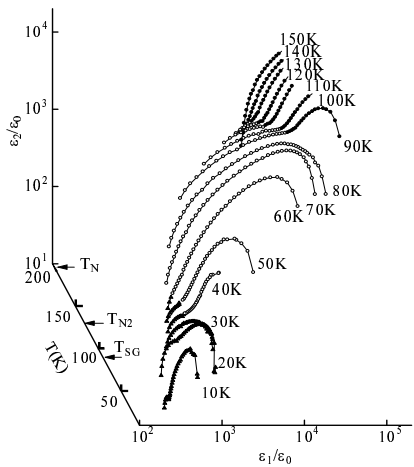}
\caption{Temperature dependence of the Cole-Cole plot. 
Three relaxations are plotted with different marks.}
\label{colecole}
\end{figure}

Reflecting the successive phase transitions, 
the dielectric response is quite complicated.
Figure \ref{f3} shows $\varepsilon_1$ of
Sr$_2$Cu$_2$CoO$_2$S$_2$ as a function of frequency.
Let us begin with the data at 10 K.
$\varepsilon_1$ shows a step-like decrease with increasing frequency, 
which is a dielectric relaxation.
With increasing temperature,
additional tails appear in $\varepsilon_1$ at low frequencies above 
40 K and 80 K,
which means that another relaxation process begins to overlap.
The tail extends to higher frequencies with increasing temperature,
and low-frequency $\varepsilon_1$ increases.
Dielectric relaxation is more clearly seen in the so-called Cole-Cole plot 
for various temperatures (figure \ref{colecole}).
in which the imaginary part of dielectric constant 
($\varepsilon _2$) is plotted as a function of $\varepsilon_1$.
Here we calculated $\varepsilon_2$ from $\sigma$ as 
$
\varepsilon_2/\varepsilon_0 = (\sigma-\sigma_{dc})/\omega\varepsilon_0,
$ 
where $\sigma _{dc}$ is the dc conductivity.
At 10 K, the data draws a single distorted semicircle,
indicating the dielectric relaxation.
With increasing temperature, 
additional relaxations begin to overlap, 
and the data draw two semicircles.
Note that the three relaxations are related to the magnetic transitions.
The lowest-temperature relaxation corresponds to
the hysteresis region from 10 to 50 K.
The second relaxation corresponds to the second antiferromagnetic 
phase below 120 K ($\sim T_{N2}$),
and the highest-temperature relaxation corresponds to the first 
antiferromagnetic phase below $T_N$.

Let us discuss the origin of the dielectric relaxation.
It should be emphasized that 
the observed relaxation is unique, 
and at least three relaxations 
appear with various magnetic phases.
Since we used polycrystalline samples, we cannot exclude 
a possibility that this anomaly comes from grain boundaries.
As Lunkenheimer \textit{et al.} \cite{lunkenheimer} have pointed out, 
anomalously large $\varepsilon_1$ at low frequencies might be
related to a depletion layer at the grain boundaries, 
what is called ``leaky capacitor'' effect \cite{he}.
However, $\varepsilon_1$ in the low-frequency limit is 
almost independent of temperature in the leaky capacitor,
whose temperature and frequency dependences are 
different from our data.
Additionally, the coincidence between the dielectric relaxations 
and magnetic phase transitions is too orderly. 

As mentioned above, we propose a self-organization
or phase-separation of holes takes place in Sr$_2$Cu$_2$CoO$_2$S$_2$.
According to this picture, the charge density is inhomogeneous in the sample, 
which could be vibrated by an external ac field 
to give large dielectric response.
Below $T_N$, the holes are gradually confined in the percolation network
and the spin-rich region (highly resistive) increases in volume.
Thus the electronic network is near the percolation threshold near $T_N$,
and goes far away from it with decreasing temperature.
The dielectric constant of percolation systems consisting of metal 
and insulator nano-composites has been extensively investigated. 
Recently Pakhomov \textit{et al.} have reported that
it diverges at low frequencies near the percolation threshold \cite{pakhomov},
whose spectrum is very similar to the 200 K data in Fig. 3.
At low temperatures, a sample away from the threshold in the 
insulating side shows a flat dielectric constant 
\cite{laibowitz},
which is also very similar to 
the data in Fig. 3.
For more quantitative analysis, 
we need the same measurement with 
single crystal samples.

In summary, we have measured the resistivity, thermopower,
dielectric constant and ac conductivity of 
polycrystalline samples of Sr$_2$Cu$_2$CoO$_2$S$_2$. 
All the transport parameters show 
characteristic change below $T_N$, 
which clearly indicates that
the charge transport is highly correlated to the spin state.
In particular, the qualitatively different dielectric relaxations 
are observed with decreasing temperature,
which well correspond to the different magnetic phases. 
We have proposed that 
the transport properties are consistently understood 
in terms of the spin-carrier phase separation below $T_N$.

\begin{acknowledgments}
The neutron powder diffraction using the high efficiency and 
high resolution measurements~\cite{hermes} were performed under the 
inter-university cooperative research program of the Institute for 
Materials Research, Tohoku University. 
This work was partially supported by MEXT, the Grant-in-Aid for 
Scientific Research (C), 2001, no.13640374.
\end{acknowledgments}



\end{document}